# ENHANCING DIGITAL FORENSICS READINESS IN BIG DATA WIRELESS MEDICAL NETWORKS: A SECURE DECENTRALISED FRAMEWORK.

Cephas Mpungu[1] Carlisle George [2] and Glenford Mapp[3]

Department of Computer Science, Middlesex University, London, UK
cm1677@live.mdx.ac.uk
c.george@mdx.ac.uk
g.mapp@mdx.ac.uk

## ABSTRACT

*Wireless medical networks are pivotal for chronic disease management, yet the sensitive Big Data they generate presents administration challenges and cyber vulnerability. This Big Data is valuable within both healthcare and legal contexts, serving as a resource for investigating medical malpractice, civil cases, criminal activities, and network-related incidents. However, the rapid evolution of network technologies and data creates complexities in digital forensics investigations and audits. To address these issues, this paper proposes a secure decentralised framework aimed at bolstering digital forensics readiness (DFR) in Big Data wireless medical networks by identifying security threats, complexities, and gaps in current research efforts. By improving the network's resilience to cyber threats and aiding in medical malpractice investigations, this framework significantly advances digital forensics, wireless networks, and healthcare. It enhances digital forensics readiness, incident response, and the management of medical malpractice incidents in Big Data wireless medical networks. A real-world scenario-based evaluation demonstrated the framework's effectiveness in improving forensic readiness and response capabilities, validating its practical applicability and impact. A comparison of the proposed framework with existing frameworks concluded that it is an advancement in framework design for DFR, especially in regard to Big Data processing, decentralised DFR storage and scalability.*

## 1. INTRODUCTION

Security threats, especially data breaches, have dramatically increased in the healthcare sector over the past few years [1]. As of August 2023, the sectors with the highest number of disclosed breaches worldwide were; healthcare, followed by education and the public sector [2]. These threats, coupled with the sensitivity of healthcare data, medical malpractices, data protection legislation requirements regarding data breaches [3], and the challenges of handling Big Data, have intensified the need for digital forensics readiness (DFR) within healthcare wireless medical networks (WMNs).

DFR is a derivative of the term digital forensics (DF). DF is a process within forensic science necessitating the identification, collection, analysis, and preservation of potential evidential data in a legally admissible manner [4, 5]. DFR is a structured configuration of systems and mechanisms in a network of digital assets that simplifies DF investigations through the proactive collection of potential evidential data [5-8] This should be accomplished at minimal cost and interruption to the business operations of an organisation as laid out in its Emergency Operations (EO) plan. DFR also enforces legal/regulatory compliance and hardens the security structures of an organisation's network [3]. Additionally, it demonstrates a strong commitment to best practices like ISO standards for DFR and security which enhances the reputation of an organisation.

In regard to the above, some researchers have proposed DFR frameworks for WMNs [6-8]. However, their proposals have recommended frameworks that handle mainly structured evidential data. The frameworks are also underpinned by centralised logging mechanisms that may create single points of failure and consequentially compromise the Confidentiality, Integrity and Availability (CIA triad) of evidential data. Other researchers like Mpungu et al. [4] proposed a framework that leverages blockchain technology to enforce some level of decentralisation but again the emphasis was on logging structured evidential data.

Furthermore, it is worth noting that the research studies cited above have also not adequately addressed a comprehensive DFR and Incident Response (IR) mechanism within healthcare Big Data networks. This void is particularly significant because, in addition to network security threats, other incidents such as medical malpractice and criminal behaviour also generate digital footprints (in the form of structured, semi-structured and unstructured data [9]) that require proactive collection. Therefore a forensically ready WMN with a secure framework that harnesses the full capabilities of Big Data (including semi-structured and unstructured data) can be extremely resourceful. This is because it can aid in the investigation of network-related incidents/threats and other healthcare-related incidents e.g. medical malpractice and criminal acts such as the Lucy Letsby infant murders at the Countess of Chester Hospital in the UK [10].

WMNs are rapidly advancing and producing substantial amounts of Big Data, including structured, semi-structured, and unstructured data [9]. This proliferation of data amplifies the potential for security incidents and also complicates the process of identifying and extracting potential evidential data [11]. Additionally, some researchers have highlighted security concerns about the vulnerability of medical devices to hacking attacks [12]. lastly, WMNs are also notably more vulnerable to attacks compared to wired networks because they use radio frequency (RF) signals for communication [13, 14].

In light of the above, this research paper emphasises that beyond network-related incidents, other healthcare-related incidents such as medical malpractice and criminal acts (e.g. the Lucy Letsby infant murders) could be promptly identified, thoroughly investigated, and effectively addressed before they escalate. Therefore a secure Big Data WMN DFR mechanism that leverages a holistic identification, collection, normalisation, storage and analysis of data (structured, semi-structured and unstructured) from disparate sources is key for ensuring effective incident response as well as the reconstruction of solid incident scenarios during digital investigations.

This paper investigates previous research within the field of DFR for WMNs and proposes and implements a novel secure decentralised DFR framework for Big Data WMNs. The paper, therefore, addresses the challenges associated with centralised logging and the complexities of handling Big Data within a WMN ecosystem to achieve an improved novel DFR state. It proposes and implements a conceptual framework that addresses Big Data WMN DFR and subsequent IR. The rest of this paper is structured as follows: Section 2 reviews related literature, Section 3 identifies framework requirements, Section 4 explores the conceptual design and orchestration system of the proposed framework, and Section 5 discusses a scenario-based case study evaluation. The paper then concludes in Section 6.

## 2. RELATED WORK

This section discusses related previous studies within WMNs that have provided foundational research as well as helped in identifying research gaps.

Several patients suffering from chronic diseases rely on WMN devices such as implantable devices or glucose monitoring devices [7]. Examples of Chronic diseases include diabetes, heart disease and cancer [15]. Under the assumption that these devices failed or were compromised by hackers as demonstrated by Radcliffe [12] and Halperin et al. [16], a digital investigation would most likely be commissioned by the relevant authorities [7].

In light of the aforementioned vulnerability of use and abuse by unauthorised users [12, 16], Cusack and Kyaw [6] proposed a DFR system for WMNs comprising security enforcement abilities as well as a capability to investigate post-events. Having conducted an in-depth analysis of various wireless medical technology architectures, focusing on their security risks, they suggested the enhancement of an existing hospital information system's WMN by introducing drones and a forensic server. The impact of their work on the field and this research is: *i)* the need *to ensure WMNs are prepared for digital forensics, it's essential to deconstruct and assess them according to industry risk criteria,* and *ii) the imperative to consider the risk of security failures/incidents within WMNs whilst developing a risk management plan.*

Building on the work done by [6], other researchers like Rahman et al. [8] also discussed DFR in WMNs focusing on Wireless Body Area Networks (WBAN). They utilised the concept of Practical Impact Analysis (PIA) to analyse potential WBAN vulnerabilities and threats. Based on their findings, they proposed solutions for a forensically ready system. Their work contributed to the field of DFR by highlighting issues associated with security vulnerabilities of WBANs (at the time) and Internet of Medical Things (IoMT) devices (in this day and age). The researchers stressed the need for the addition of an *isolated WBAN secure architecture*. Their proposal helped inform this research in the area of *wireless network segmentation* for purposes of *enforcing security*.

Cusack and Kyaw [6] later teamed up with another researcher named Lutui [7]. They proposed a modified DFR framework for wireless medical devices (WMedSys) based on their previous work in [6]. The proposed framework aimed to streamline and reduce the time and cost of digital forensics investigations (DFIs). The conceptual design consists of a Kali Linux Pi-drone that scans and captures wireless signals sending them to a wireless forensic server(WFS). Other components of this framework are an Intrusion Detection System (IDS), integrity Hashing Server, Centralised Syslog server (SPLUNK), Wireless Access Point (WAP), Remote Authentication Dial-In Service (RADIUS) server and a web server. The evaluation was done using a thematic expert analysis in NVIVO. The research made a novel contribution to the field of digital forensics and impacted the current work undertaken by suggesting a *secure*, *cost-effective, time-effective* streamlined mechanism for the capture of potential (structured) evidential logs whilst *enforcing evidential data integrity*.

The core reason behind conducting this research is twofold: first, there has been a significant rise in security breaches within the healthcare industry, and second, the inadequate utilisation of Big Data to attain a comprehensive digital forensics readiness environment has created gaps that need to be addressed. Three major gaps were identified in the reviewed literature:

1) *A secure decentralised DFR logging/storage system in WMNs has not been implemented.*

2) *A Big Data framework for the major implementation of DFR in WMNs to accommodate all healthcare-related incidents has not been proposed or implemented.*

3) *There is a need to establish a secure investigative environment that accommodates all relevant parties, including forensic investigators.*

In light of the identified gaps, it is pertinent to highlight that WMNs and their affiliated devices constantly generate (sensitive) large amounts of data (Big Data) as they evolve which could be the target of ransomware attacks and data breaches [17]. The Big Data generated by a WMN ecosystem is characterised by the five major Vs (Volume, Velocity, *Variety* (structured, semi-structured and unstructured), Veracity and *Value*) [18].

This abundance of data amplifies the risk of data mismanagement and provides a larger surface area for malicious actors to exploit through hacking, ransomware attacks, supply chain breaches, phishing, and other harmful activities. An example of such an attack is the 2017 Wannacry ransomware attack on the UK's National Health Service (NHS), which affected numerous computers across the globe due to vulnerabilities in the organisation's systems, including a known flaw in Microsoft's Windows operating system [19]. Furthermore, it is challenging to detect fraudulent or malicious activities within this vast pool of Big Data, as highlighted by Zawoad and Hasan [18]. Hence effective and secure management of Big Data DFR is essential to mitigate these risks in WMNs. The focus of this research is to harness effective DFR and incidence response from the *variety* and *value* attributes of Big Data generated by WMNs.

Hence this *variety* of *valuable* Big Data can also be securely and systematically stored in a decentralised configuration and leveraged for Digital forensic intelligence and investigations. This research paper will demonstrate how the use of structured, semi-structured, and unstructured evidential data can be highly valuable in mitigating and investigating other healthcare-related incidents, such as medical malpractice and criminal offences, in addition to the network security incidents discussed within the reviewed research.

Big Data is already being leveraged by most business sectors, including healthcare, to derive meaningful and actionable business intelligence insights [20]. In that regard, this paper proposes a similar modified configuration mechanism that will be used to derive actionable Digital Forensics Intelligence (*DFIntel*) from healthcare-generated Big Data. Tableau [21] defines Business intelligence (BI) as a combination of "*business analytics, data mining, data visualisation, data tools and infrastructure, and best practices to help organisations to make more data-driven decisions.*"

*DFIntel*, in context to this paper, is a collection of actionable digital forensics intelligence derived from the use of digital forensic tools and infrastructure deployed for Big Data mining, digital forensics imaging, analytics, and visualisation, underpinned by best practices to enable organisations to achieve effective DFIs.

In conclusion, the motivation for this research paper was initially derived from the dramatic increase in security incidents within the healthcare sector. The work was then inspired by the work done by [6] [7] and [8]. The researchers highlighted the security risks associated with WMNs (at that time), the need for DFR within WMNs, and proposed frameworks to address these issues as expounded in the literature review. This research paper builds on their work with the intent of resolving the highlighted identified gaps to facilitate: *1) decentralised evidential data log storage 2) Big Data DFR within WMNs and 3) a secure investigative environment that accommodates all relevant parties*. Lastly, a comparison of the existing WMN DFR frameworks discussed above and the proposed framework in the current research is presented in Table 4, Section 6.

## 3. REQUIREMENTS FOR THE PROPOSED FRAMEWORK

The ISO/IEC's 27043:2015 incident investigation planning process guided the identification of requirements for the proposed framework [22]. The process comprises the following six phases explained below:

### 3.1. Definition of scenarios

Defining scenarios is crucial for the conceptualisation of simulations of incidents that could initiate a digital forensic investigation or audit of the WMN. A risk assessment should be conducted to identify potential vulnerabilities and threats within the WMN. Scenarios should be then defined based on the risk assessment analysis and also based on the current threats and incidents facing WMNs and healthcare in general. Scenarios can include past ransomware attacks on the NHS, internal data breaches, DDOS attacks, medical malpractices and healthcare-related criminal offences. Identification of scenarios enlightens the next phase (identification of potential data sources) of the planning progress group.

### 3.2. Identification of potential evidential data sources

This phase is underpinned by the scenario-definition process and involves the explicit identification of all evidential data sources within the WMN. This entails proper visibility of the network especially critical network assets to determine what kinds of data need to be logged for effective DFR. A detailed mapping of the network using relevant network mapping tools should be done to support this.

### 3.3. Pre-incident planning, gathering, storage and handling of potential evidential data

The proposed framework should clearly define pre-incident planning, as well as the proactive gathering, storage, and handling of potential evidential data, including structured, semi-structured, and unstructured data. Appropriate planning and generally acceptable standards should be applied and the data stored in a secure and admissible form for purposes of analysis and effective incident response.

### 3.4. Pre-incident analysis of potential evidential data planning

This planning phase should be consistent with the kind of data that is going to be logged for purposes of DFR. The analysis should, therefore, be planned considering factors like the network throughput, analysis tools/software, expertise, remote analysis capabilities, notifications configuration, and false positives minimisation, amongst others.

### 3.5. Incident detection planning

This phase involves establishing protocols, tools and a secure environment to identify and respond to potential security breaches or unauthorised activities promptly. Examples can include setting up intrusion detection systems (IDS), SIEM-configured alert mechanisms, and leveraging appropriate Big Data framework capabilities for processing and analysing large volumes of log and event data in real-time.

### 3.6. System architecture definition

This phase focuses on identifying and documenting all hardware, software, network components, and their interactions. These include hardware, software, network topology, data flow, access controls, logging and monitoring, secure data storage, interactions, and critical systems. This phase is important because it enables investigators and other relevant third parties to quickly locate and access relevant data sources during investigations.

## 4. DESIGN OF THE PROPOSED WMN DFR FRAMEWORK

The novel WMN DFR framework that was developed for this research consists of three main configuration benchmarks: (a) WMN virtualisation, (b) decentralised storage, and (c) secure authentication (for DFR environment access). These configurations are also the implementation platforms for the *secure proactive collection of potential evidential logs* to ensure efficient DFI, network audits, regulatory compliance and business recovery. Network mapping tools like SolarWinds Network Performance Monitor or Nmap are used to gain profound visibility of all the network assets that need to be secured and logged. The network design of the novel WMN DFR framework is shown in Figure 1 below. The red arrows accentuate the flow of DFR data (DFIntel) and the green oval segment highlights the Big Data DFR module (*BdDFR*).

**Note:** *The picture resolution of the configuration segment depicting **Digital Forensics Readiness** in the network diagrams (Figure 1 and subsequent figures) was increased to highlight the focus and novelty of this research. Meanwhile, the surrounding configurations were intentionally set to a lower resolution. All the figure network diagrams used in this section were developed by the research authors.*

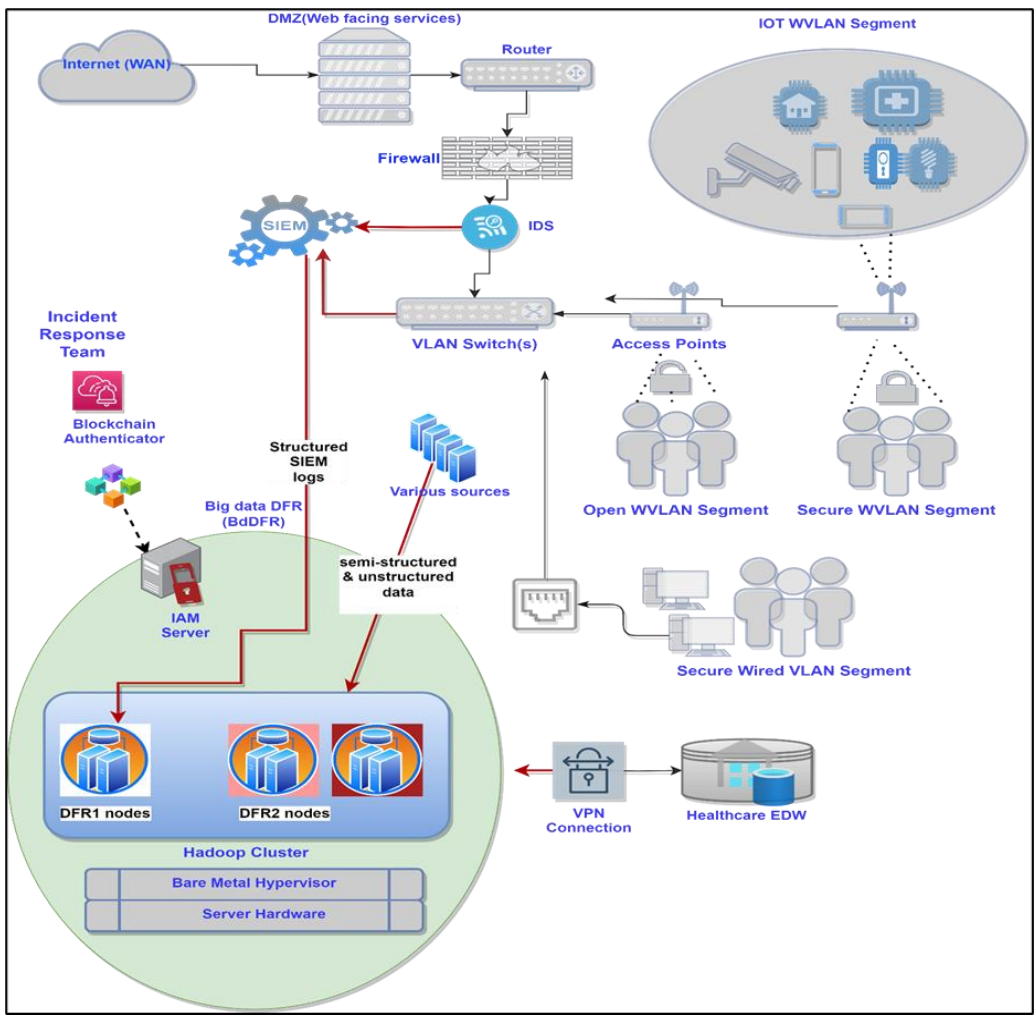

Figure 1. WMN DFR framework overview

Table 1 below denotes the full forms of the acronyms used in Figure 1 (WMN DFR framework overview). Table 2 highlights the Incident Response Team (IRT), their roles, classes and levels.

Table 1 Figure 1 Acronyms full forms

| Acronym | Full Form |
|---|---|
| DMZ | Demilitarized zone |
| VLAN | Virtual Local Area Network |
| WVLAN | Wireless Virtual Local Area Network |
| IOT | Internet of Things |
| EDW | Enterprise Data Warehouse |
| IDS | Intrusion Detection System |
| SIEM | Security Information and Event Management |
| IAM | Identity and Access Management (server) |
| VPN | Virtual Private Network |

Table 2 Incident Response Team (IRT)

| Title | Role | Class | Level |
|---|---|---|---|
| IRT Leader | Overall incident response coordination, strategy, and communication. | B | 2nd responder |
| Incident Manager | Assist the IRT Leader in managing incidents and serves as a backup in their absence. | B | 2nd responder |
| Technical Lead (Forensics Specialist) | Leads the technical aspects of incident response, digital forensics, and data recovery. | B | 2nd responder |
| Network Security Specialists | Day-to-day network monitoring, troubleshooting and implementation of security controls. Collaborates with the Technical Lead during incidents. | A | 1st responder |
| System Administrator | Day-to-day management and security of the organisation's systems, servers, and devices. Coordinates with the Technical Lead during incidents. | A | 1st responder |
| Data Privacy Officer (DPO) | Ensure compliance with data protection and privacy regulations. Assesses the impact of incidents on patient and employee data privacy | C | 3rd responder |

|  | during incidents. |  |  |
|---|---|---|---|
| Legal Counsel | Provides legal guidance and ensures that incident response efforts comply with legal requirements. | C | 3rd responder |
| HR Representative | Plays a critical role in incident response, particularly in incidents that involve employee misconduct, insider threats, or breaches of employee data. | C | 3rd responder |

### 4.1. WMN Virtualisation

The WMN is virtualised by segmenting it into isolated virtual networks (subnets) using VLAN switches as indicated in Figure 1. Each subnet is configured with its own set of security controls and policies depending on its role and assets. This mitigates the spread of malware, unauthorised access and other malicious activities and also enables streamlined logging of evidential data within the BdDFR. The network segmentations include the *IoT WVLAN segment, open WVLAN segment, secure WVLAN segment, secure wired VLAN segment, healthcare EDW, and the BdDFR.* The Hadoop cluster within the BdDFR is also virtualised and housed on a bare metal hypervisor subnet for additional security.

### 4.2. Decentralised Storage

To ensure effective decentralised logging of all structured, semi-structured and unstructured logs and data relevant to DFR, the placement and integration of the IDS, SIEM and the Hadoop Cluster nodes are crucial. The integration of the three systems is achieved as follows:

- For DFR purposes, the IDS and SIEM are positioned between the firewall and the VLAN switch. This ensures effective traffic visibility, inspection, anomaly detection, log collection, quick response, and compliance with security regulations.

- In that regard, IDS sensors are strategically deployed within the WMN to monitor network traffic and identify potential intrusions in real-time.

- The IDS sensors send alerts and event data to the SIEM for centralised monitoring, correlation and storage.

- The SIEM is also configured to collect, normalise and centralise structured log data from other sources within the WMN, including servers, firewall logs, network access logs, network devices, and medical devices.

- Real-time event correlation and analysis rules are set up on the SIEM to detect security incidents and anomalies and alert the first responders (class A) of the IRT.

- The SIEM is configured to store all logs in centralised storage accessible outside the DFR to allow day-to-day monitoring and interaction by class A IRT personnel. This configuration also facilitates initial response to incidents before escalations are made to class B or C IRT personnel.

- Finally, the SIEM is also configured to forward a copy of all its data to designated nodes (DFR1) within the BdDFR's Hadoop cluster. This is done to facilitate further comprehensive historical analysis and a secure DFR that may also be accessible to relevant third parties outside the WMN.

- The Hadoop cluster is the implementation mechanism for decentralised storage using HDFS (Hadoop Distributed File System). This provides a platform for both distributed storage and processing. The cluster is isolated from the WMN and configured on a private cloud to maintain a secure DFR and DFI environment.

- The Hadoop cluster is configured to store structured security-related data logs from the SIEM in DFR1 nodes. Semi-structured and unstructured data (such as emails, CCTV footage, medical scans, IoMT data, radiography reports, social media posts) are stored within DFR2 nodes.

The BdDFR module is extracted from Figure 1, depicted in Figure 2 below, and further elaborated upon in the subsequent paragraphs.

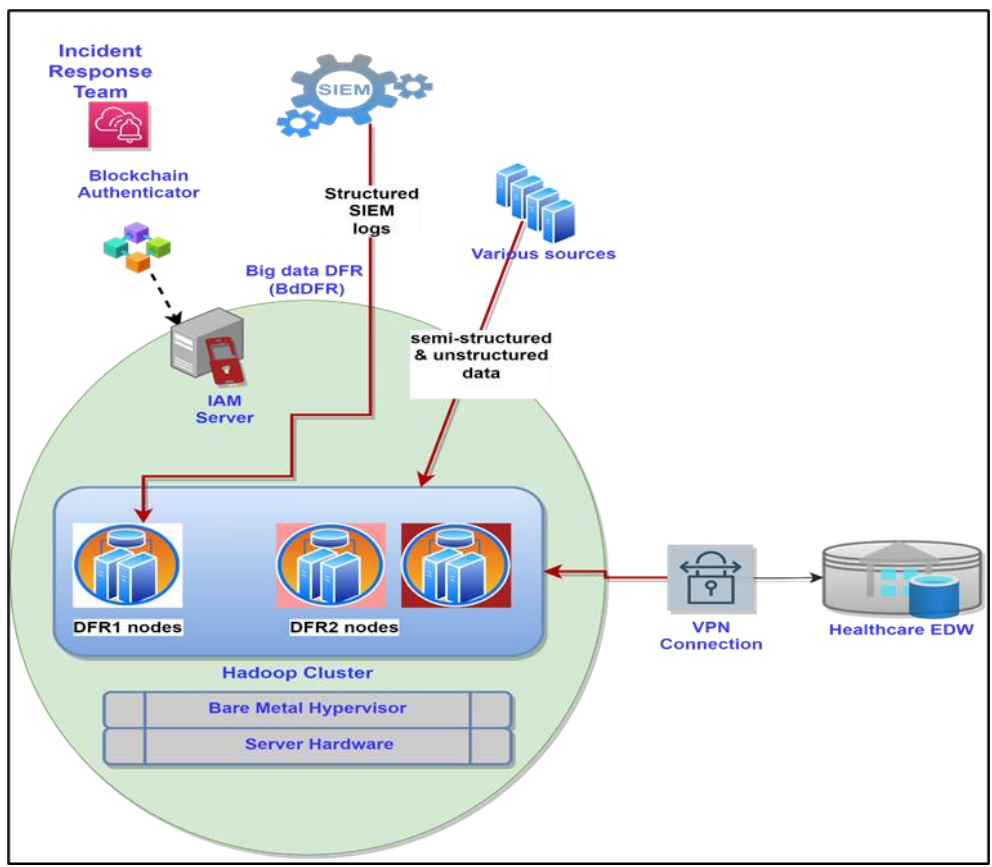

Figure 2. Big Data DFR module (BdDFR)

The SIEM, located outside BdDFR, is the primary point/first point of investigation for digital forensics alerts and investigations by class A IRT. The enforcement of regulatory compliance is also primarily achieved here.

However, one major weakness of SIEMs is their inability to scale or handle varying loads of Big Data [23]. SIEM-related logs can also suffice for network security incident investigations but would not fully support other healthcare-related incident investigations like medical malpractices and some criminal offences. It was for this reason that this paper proposed the addition of a Big Data DFR module known as *BdDFR*. This comprises a decentralised Hadoop cluster partitioned into DFR1 and DFR2 nodes. DFR1 nodes are configured for storage and historical analysis of structured security-related logs generated by the SIEM. DFR2 nodes are configured to ingest semi-structured and unstructured logs as well as data from various (relevant) sources within the WMN that cannot be processed by conventional SIEMS and yet are critical for digital forensics investigations. The justification of this framework setup, from a context of a Big Data WMN, is that *DFR should be configured considering that some incidents and subsequent investigations will require much more than just SIEM security logs.*

The Hadoop cluster is integrated with the healthcare EDW using a secure VPN connection. The benefits of this integration are vast for most organisations [24]. However, in the context of the current work, it would enhance the holistic DFR and IR approach and provide more context to investigations.

The healthcare EDW, located outside the BdDFR, contains essential structured data for BI but may also support investigations if needed. Notably, the main DFR modules for this research are the SIEM and the Hadoop ecosystem. The assumption is that the healthcare EDW is already in place, as most organisations handling Big Data use EDWs for BI.

Time synchronisation is crucial for accurate metadata timestamps, event correlation, effective investigations, and regulatory compliance. Therefore, all BdDFR assets are configured to synchronise with the WMN's NTP (Network Time Protocol) server.

Regarding user data storage, the framework design caters to three main users of the WMN namely: *Day-to-day users, network IT personnel and the IRT*. Day-to-day users comprise authorised and unauthorised users. Authorised users include healthcare employees, patients, visitors, and connected devices. Unauthorised users include external threat actors, internal threat actors, and rogue devices. Network IT personnel (class A IRT) play a critical role in maintaining the WMN's functionality and security by ensuring its robustness, troubleshooting, access control, updates and security audits. These professionals are part of the IRT due to their expertise in managing and mitigating network-related incidents. However, most of their day-to-day work remains confined to the exterior of the BdDFR as they would have access to all the relevant assets and resources needed for their roles. This helps maintain the integrity and authenticity of the data and logs captured within the BdDFR environment.

The aforementioned users generate substantial volumes of diverse data, constituting structured, semi-structured, and unstructured data, which can serve as potential sources of evidentiary information. Examples of structured evidential data generated from these users include network traffic, successful logins, failed logins, detected attack activity, inconsistent bandwidth levels, errors, failures, access events, end-user application logs, user account changes e.g. deletion, password changes, object access denials, large file transfers, and attempted admin access. This is configured, monitored and analysed at the IDS and SIEM and then logged at both the SIEM storage and DFR1 nodes. Examples of semi-structured evidential data include electronic health records (EHRs) in XML or JSON format, email, excel entries, and medical imaging data in DICOM format. Examples of unstructured evidential data examples include email content, raw CCTV footage, radiography reports, social media posts, medical research, PDF documents, audio recordings, and patient notes. Semi-structured and unstructured evidential data is logged in DFR2 nodes.

Minimisation of noise and false positives is configured at the SIEM by first establishing a baseline normal behaviour for the WMN and its associated systems. Based on this, alert thresholds, correlation rules, User and Entity Behaviour Analytics (UEBA), whitelisting, and blacklisting are then implemented.

To ensure the confidentiality, integrity, and availability of potential evidential data and all other data processed within the Wireless Medical Network, encryption measures are implemented on communication channels (data in transit) and data at rest. Communication channels are secured using TLS (Transport Layer Security). Data at rest is secured using a combination of wireless network security best practices. These comprise regular software updates, antivirus software, encryption (disk-level and file level), ACLs, strong password policies, disabling unused ports, and physical security of the WMN assets. Additionally, all the encryption keys are stored in a secure location and regularly rotated to impede unauthorised access.

Data Anonymisation is enforced within DFR1 and DFR2 to ensure compliance with legislation governing the security of PII (Personally Identifiable Information) and special category data. It entails removing or obscuring any identifying information that could be used to link the data back to an individual, thus rendering the data anonymous. Methods like masking, aggregation, data perturbation and cryptographic techniques can be used to obscure this kind of data [25]. Furthermore, data control mechanisms should be put in place to implement data retention policies to mitigate the risk of data leaks.

### 4.2.1. Digital Forensics Readiness Configuration

As earlier discussed, the initial *readiness* of potential evidential data is configured and enforced primarily at the SIEM interface and its storage server (accessible outside the BdDFR). This configuration is one of the features that make this framework scalable even for a wireless network handling small data (minimal data). However, on a Big Data WMN, the synergy of the SIEM and BdDFR is paramount. Alerts triggered by the SIEM are first investigated by class A IRT and escalations to class B or C IRT will only be made depending on the severity of the incident. This mitigates false positives and maintains a sanitised BdDFR environment. Alerts triggered by the BdDFR are considered high priority and should be sent to both class A and class B IRT by SMS and email. Involvement of class C IRT should only be triggered if the incident severity warrants this.

### 4.2.2. Big Data (DFR) Environment (BdDFR)

The Big Data environment comprises a Hadoop cluster hosting DFR1 and DFR2 nodes. The nodes are configured as a *decentralised* repository for the storage of Big Data (in form of structured, semi-structured and unstructured data) from a variety of sources. Apache Hadoop consists of a distributed architecture, with multiple nodes that work simultaneously to store and process large amounts of data. Its two major components are HDFS and MapReduce. HDFS is a *decentralised* file system that stores data across multiple nodes in a cluster. It supports *high-speed* access to large files and provides *fault tolerance* by replicating data across multiple nodes. These highlighted features of HDFS make it ideal for capturing part of the *novelty* requirements of this paper in light of processing Big Data for purposes of DFR and IR within a decentralised storage platform that supports business continuity (in case of security incidents).

Hadoop also enables support for multiple interfaces to allow third-party access. Hence applications can securely access Hadoop data and services through various APIs and protocols. It provides multiple interfaces, including HDFS, MapReduce, machine learning plugins, Apache spark, Hive, Pig, HBase, Hadoop Web Graphical User Interface (GUI), Command Line Interface (CLI), Hue, JobTracker UI, and ResourceManager UI (YARN) [26]), which enable developers to build applications for processing large data sets, machine learning, forensic analytics, and predictive analytics. This subsequently fosters interoperability between Hadoop and other systems, allowing data to be shared and processed across different systems ultimately allowing third parties secure access to BdDFR (e.g. investigators, data scientists, and healthcare researchers among others from different organisations).

## 4.3. Secure Authentication

The WMN is virtualised by segmenting it into isolated virtual networks (subnets) using VLAN switches as indicated in Figure 1. Each subnet is configured with its own set of security controls and policies depending on its role and assets. This mitigates the spread of malware, unauthorised access and other malicious activities and also enables streamlined logging of evidential data within the BdDFR. The network segmentations include the *IoT WVLAN segment, open WVLAN segment, secure WVLAN segment, secure wired VLAN segment, healthcare EDW, and the BdDFR*. The Hadoop cluster within the BdDFR is also virtualised and housed on a bare metal hypervisor subnet for additional security. Authentication to the BdDFR is managed using IAM for internal users, and remote users are authenticated with an additional layer of blockchain-based authentication

The configuration of the BdDFR described in the previous sections is illustrated in Figure 3

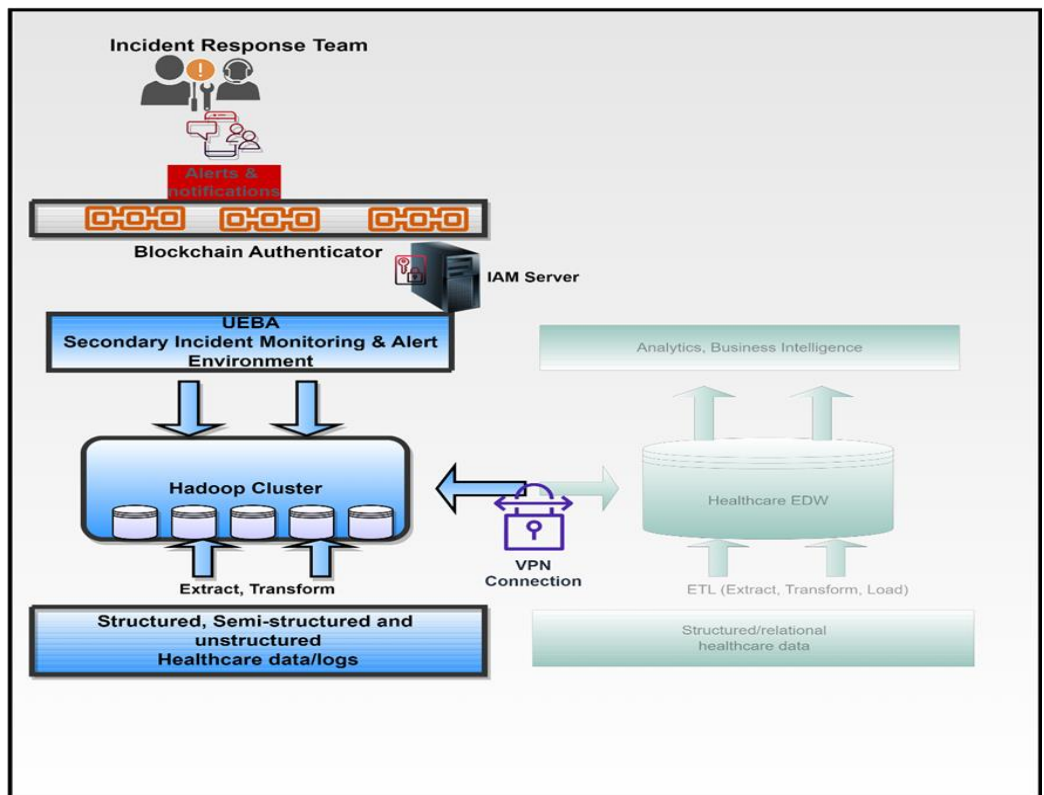

Figure 3. BdDFR configuration.

## 4.4. Leveraging Big Data for DFR

To offer a clearer understanding of how the proposed framework leverages Big Data for DFR, Table 3 below shows the types of data/logs essential for DFR and investigations related to specific scenarios. These include *patient insurance claims, cases of medical negligence leading to fatalities, malfunctions of IoMT devices, incorrect diagnosis, and instances of employee misconduct*. The data/logs are captured within the DFR1 and DFR2 nodes of the Hadoop cluster depending on their structure.

Table 3 Evidential logs.

| Scenario | Structured data/logs (DFR1) | Semi-structured data/logs (DFR2) | Unstructured data/logs (DFR2) |
|---|---|---|---|
| Insurance claims | Billing, claims and EHR data, medical reports, and related metadata. | Email and chat logs | Social media posts |
| Medical negligence | EHR data, authentication logs, patient admission and discharge records, and staff access logs. | Email and chat logs, physician diagnostic notes. | Xray scans, CT scans, CCTV footage, video/audio recordings of medical procedures |
| IoMT malfunctions | Pcap files, device maintenance and calibration data, device status and error logs, and update logs. | sensor data (depending on the IoMT device) | Sensor data, and surveillance camera footage(to rule out intentional sabotage). Maintenance reports |
| Incorrect diagnoses | EHR data, medical reports, patient admission and discharge records. | Email and chat logs, physician diagnostic notes. | Medical procedures Video/audio recordings, CCTV footage. |
| Employee misconduct | Access logs indicating staff entry and exit times | Email and chat logs | CCTV footage and social media posts. |

## 4.5. Incident Response Configuration

The Hadoop cluster functions as a repository for different data types, including structured logs from the SIEM, along with semi-structured and unstructured logs and data from additional sources. This data consolidation facilitates long-term analysis and enables the detection of intricate security incidents as well as medical malpractices that might elude the monitoring capabilities of structured SIEM logs. Concurrently, a UEBA (User and Entity Behaviour Analytics) system conducts a comprehensive analysis of all the data logs stored within the Hadoop cluster nodes. UEBA leverages advanced analytics and machine learning techniques to pinpoint deviations and patterns in behaviour, subsequently prioritizing alerts according to potential risk levels. It excels in identifying insider threats and other anomalous activities within the realm of semi-structured and unstructured data and logs, areas where traditional SIEMs may struggle to process or may inadvertently overlook.

This integrated approach ensures a robust DFR and IR. Alerts from both the SIEM and UEBA systems guide security teams in responding to potential incidents promptly. The IRT first responders can then assess these alerts, validating or dismissing them, and take appropriate actions as needed. The historical data archived in the Hadoop cluster also becomes a valuable resource for retrospective analysis, forensic investigations, and the continual improvement of incident response procedures. For imaging of the evidential data from the relevant BdDFR nodes, a write blocker (to ensure read-only access and preserve data integrity) is attached and the image is extracted using digital forensic software. To ensure fast and secure investigations, it is possible to allocate a dedicated node within the Hadoop cluster specifically for imaging and investigative tasks.

# 5. A SCENARIO-BASED EVALUATION OF BdDFR (*THE LUCY LETSBY CASE*)

The proposed framework was evaluated by analysing a real-world incident involving the utilisation of structured, semi-structured, and unstructured logs within a Big Data WMN for effective alerting and investigation as detailed below.

The Lucy Letby case concerns a nurse in the United Kingdom who was arrested and charged with multiple counts of murder and attempted murder of infants in her care at the Countess of Chester Hospital neonatal unit [10]. The murders are believed to have begun in *2015* and were detected in *2018* when an investigation was initiated. In the context of DFR, BdDFR, utilising UEBA analytics, would have played a crucial role in uncovering any suspicious activities, patterns, or discrepancies in a timely manner and alerted the relevant IRT as depicted in Figure 4 below and explained in the subsequent sections.

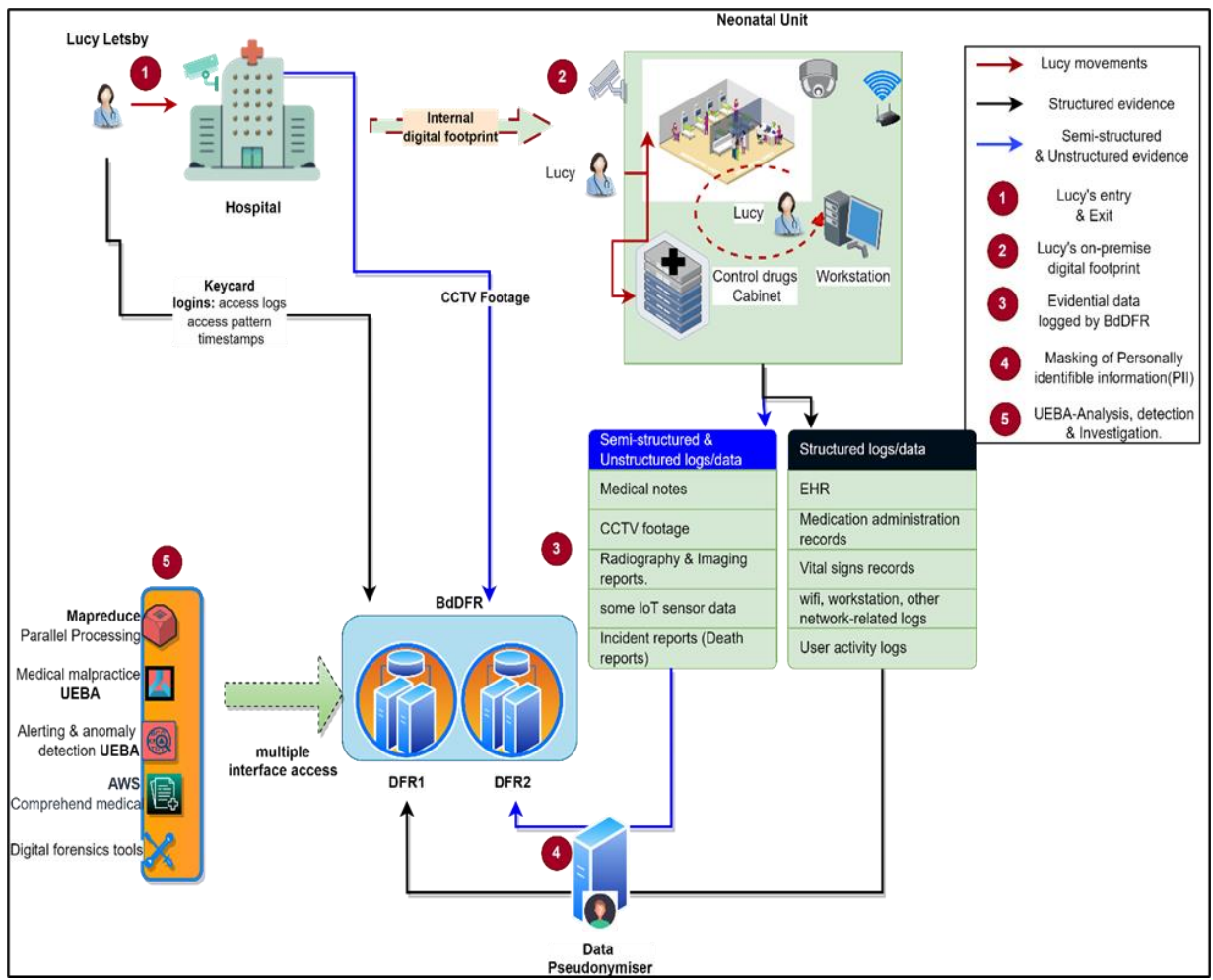

Figure 4 Early detection of the Lucy murders using BdDFR

## 5.1. Access patterns

BdDFR's UEBA system would monitor Lucy's access to restricted areas, medication rooms' control drugs cabinets and medical equipment as shown in Units 1 and 2 of Figure 4. Any unusual or unauthorised access attempts, especially during inappropriate hours or in excessive frequency, would have been flagged for investigation.

## 5.2. Medication administration

BdDFR's UEBA would analyse logs related to medication administration captured from Unit 2 of the neonatal unit, comparing Lucy's actions to standard practices within the known baseline. Any abnormal patterns, such as excessive doses, unusual medication choices, or discrepancies in medication administration timings, could have raised red flags.

## 5.3. Device interactions

The system would monitor Lucy's interactions with medical devices or equipment in Unit 2. If there were instances where she accessed or tampered with devices inappropriately or in a manner inconsistent with her role or her cohorts, such actions could have been identified as suspicious.

By detecting these anomalies and raising alerts, BdDFR would have provided insights to the IRT, enabling them to investigate and address concerns early on. It could have potentially triggered a thorough examination of Lucy Letby's actions, leading to a faster discovery and prevention of the escalation of infant deaths. The data pseudonymiser in Unit 4 ensures that all PII is obfuscated or anonymised, safeguarding individual identities and privacy.

Having evaluated the proposed WMN DFR framework, Table 4 below presents a comparison between existing WMN DFR frameworks and the proposed framework in the current research.

Table 4. A comparison of existing and current WMN DFR Frameworks.

| WMN DFR Configuration Attributes | Existing Frameworks. | Proposed framework |
| --- | --- | --- |
| 1. DFR configuration and storage | Yes [4], [6], [7], [8] | Yes |
| 2. Effective monitoring, logging and alerting | Yes [4], [6], [7], [8] | Yes |
| 3. Decentralised storage of DFR information | Partially [4] | Yes |
| 4. Secure authentication to DFR storage | Yes [4], [6], [7], [8] | Yes |
| 5. Consideration of Big Data and its Characteristics | No | Yes |
| 6. Scalability | No | Yes |

The comparison Table 4 above concludes that the proposed framework is an advancement in the design of frameworks for DFR especially in light of the processing of Big Data, decentralised DFR storage and scalability.

## 6. CONCLUSION

In conclusion, security incidents have increased and will keep escalating within the healthcare domain due to the monetary value attached to personal healthcare data and the constant technological additions to WMNs respectively. Hence, WMN designs and configurations need to be appropriately aligned to meet DFR and IR requirements whilst handling both structured data and Big Data (including structured, semi-structured and unstructured data). Previous researchers in the domain of DFR in WMNs have proposed suitable (but) centralised frameworks for handling (majorly) structured data, but the present work focused on proposing a novel decentralised DFR framework for handling structured, semi-structured and unstructured Big Data to solve the aforementioned identified gaps. The novelty of the proposed DFR framework is captured in three aspects: *i) DFR implementation in healthcare Big Data wireless environments, ii) decentralisation of evidential data storage using HDFS and iii) secure support for multiple interfaces to allow comprehensive analysis and third-party access.* Evaluation of the framework via a real-world use case concluded that it can be effective in detecting anomalies, raising alerts and providing useful insights for further investigation.

Future work will focus on the development and evaluation of a prototype for the proposed WMN Big Data DFR framework. Additionally, an incident response Linux script with a user-friendly interface tailored for Linux-Hadoop environments will be developed to streamline digital forensic investigations. Future efforts will also include integrating the script with machine learning models to enhance its functionality and effectiveness.

## ACKNOWLEDGEMENTS

The authors extend their gratitude to Middlesex University and specifically the Computer Science Department.

**Authors**

Cephas Mpungu is a final-year PhD researcher and an Hourly Paid Lecturer in the Department of Computer Science at Middlesex University, London. He has a First Class BSc Honours degree in Information Technology and an MSc in Electronic Security and Digital Forensics with Distinction. He is also a Red Hat certified systems administrator (RHCSA). His areas of expertise include digital forensics, Linux systems administration and Cybersecurity.

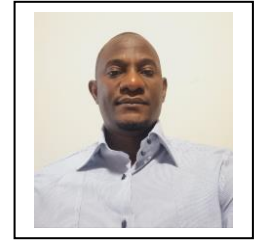

Dr Carlisle George is an Associate Professor at Middlesex University, London (UK) and a lawyer. His main areas of expertise include: Information Technology Law; Privacy and Data Protection; Intellectual Property Law; and Legal Aspects of eHealth, mHealth, Health Informatics, Digital Forensics and Data Science . He has published many academic papers on aspects of his areas of expertise and co-edited two books on eHealth and medical informatics.  He has also contributed to several European Union (EU) projects as a senior legal expert/advisor, and is involved in varied activities where his expertise is required.

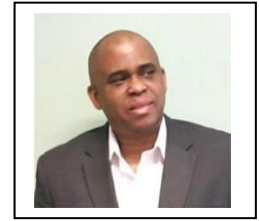

Dr Glenford Mapp is an Associate Professor at Middlesex University, London. His primary expertise is in the development of new technologies for mobile and distributed systems. Glenford does research on Y-Comm, an architecture for future mobile communications systems. He also works on service platforms, cloud computing, network addressing and transport protocols for local environments. He is currently focusing on the development of fast, portable services that can migrate or replicate to support mobile users.

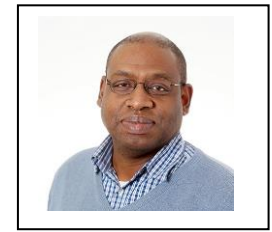